\documentclass[12pt,oneside,reqno]{amsart}

\theoremstyle{definition}

\theoremstyle{remark}

\numberwithin{equation}{section}

\usepackage{amssymb}

\begin{document}

\title{Molecular random walks and \\
invariance group of the Bogolyubov equation}

\author{Yuriy\, E.\, Kuzovlev}
\address{Donetsk Institute for Physics and Technology of NASU,
ul.\,R.\,Luxemburg 72, Donetsk 83114, Ukraine}
\email{kuzovlev@kinetic.ac.donetsk.ua}





\begin{abstract}
Statistics of molecular random walks in a fluid is considered with
the help of the Bogolyubov equation for generating functional of
distribution functions. An invariance group of solutions to this
equation as functions of the fluid density is discovered. It results
in many exact relations between probability distribution of the path
of a test particle and its irreducible correlations with the fluid.
As the consequence, significant restrictions do arise on possible
shapes of the path distribution. In particular, the hypothetical
Gaussian form of its long-range asymptotic proves to be forbidden
(even in the Boltzmann-Grad limit). Instead, a diffusive asymptotic
is allowed which possesses power-law long tail (cut off by ballistic
flight length).

\end{abstract}


\maketitle

\baselineskip 24 pt

\markboth{}{}

\section{Introduction}

Random wandering of particles of the matter is mechanism of diffusion
and many other transport processes as well as the source of various
noises and fluctuations. What kind of statistics can rule over it?
This important question never was considered in the framework of
rigorous statistical mechanics. It may seem that anyway the answer is
obvious: if in the Lorentz gas statistics of random walks is
asymptotically Gaussian \cite{sin} then all the more it must be
Gaussian in usual gas. However, such a reasoning loses its
convincingness when one notices that it indirectly equates (i)
chaotic nature of random walk in the sense of Hamiltonian dynamics
and (ii) randomness of the walk in the sense of the probability
theory. The first, as is well known, is characterized by the
``mixing'' of the system's phase trajectories \cite{kr}, whereas the
second by the ``statistical independence'' or simply ``independence''
\cite{kol} of their constituent parts. But already Krylov \cite{kr}
thoroughly explained that generally the first does not imply the
second. Therefore the question remains in force.

It can be comprehended by the example of $\,N\,$ hard balls in a box
or on torus if treating their motion as the motion of single ball in
$\,3N$-dimensional billiard \cite{gal} which is formed by
$\,N(N-1)/2\,$ convex scatterers and resembles the Lorentz gas. How
much far in time should be apart different fragments of the
trajectory of this ball in order to behave as statistically
independent one on another? It appears that not less than by the time
necessary for the ball to know about existence of all the scatterers
and thus about organization of the billiard. Let us express this
characteristic time through the mean free-flight time in the original
three-dimensional system,\, $\,\tau\sim (\pi r_0^2\nu v_0)^{-1}\,$,
where $\,r_0\,$ is diameter of the balls,\, $\,v_0\,$ is their
typical velocity,\, $\,\nu =N/\Omega\,$ their mean concentration,
and\, $\,\Omega\,$ is volume of the original system. Since the
collision rate in the $\,3N\,$-dimensional billiard is equal to
summary rate of collisions between arbitrary balls in the
three-dimensional system, approximately\, $\,N/2\tau\,$, then the
mentioned characteristic time can be estimated as\,
$\,[N(N-1)/2]/[N/2\tau ]\sim \,N\tau\sim \Omega/\pi r_0^2v_0\,$. Even
for $\,1\,$cm$^3\,$ of the air that is time on order of $\,1000$
years!

If so, then ergodicity of behavior of $\,N$-particle system can be
expected not sooner than after observing it during time intervals
$\,\gg N\tau\,$. From the point of view of real physical
many-particle systems (let even closed ones), that is quite
inaccessible time. As to reality, only more or less peculiar parts of
the system's phase trajectory are observable. Thus, it was true
remark \cite{rar} that the role of ergodicity in physics is strongly
exaggerated since there the limit $\,N\rightarrow \infty\,$ precedes
the limit $\,t/\tau \rightarrow\infty \,$.

Similar conclusions follow from the fact that the number of initial
conditions what determine trajectory of a particle of usual gas is
not $\,6\,$, as in the case of Lorentz gas, but $\,6N\,$.
Correspondingly, the trajectory can display enough its individuality
(and be recognized) only after $\,N\,$ free flights and $\,N\,$
collisions, which requires time $\,\sim\,N\tau \,$. All the more, to
disclose its ergodic properties and observe how it divides into
statistically independent parts we need in time scales $\,\gg N\tau
\,$.

To speak more specifically, the non-ergodicity of a system of
(infinitely) many particles does mean that rates of relaxation and
fluctuation processes in it are different at different phase
trajectories, as well as at diverse parts of one and the same
trajectory \cite{bk3}. For instance, individual diffusivity of a gas
particle undergoes low-frequency fluctuations \cite{i1} which
manifest themselves in a substantial difference of asymptotic
probability distribution of the particle's displacement from the
Gaussian distribution.

The theoretical tools for rigorous investigations of such a
statistics were developed by N.\,Bogolyubov \cite{bog} and his
followers. That are the Bogolyubov-Born-Green-Kirkwood-Yvon (BBGKY)
hierarchy of equations and the equivalent Bogolyubov's equation for
generating functional of many-particle distribution functions. But
nobody have learned how to use them avoiding truncations of the
hierarchy on the hypothesis about some ``obvious'' ``statistical
independence'' adopted from the probability theory. Hence, there is
necessity of methods which can treat the basic equations
respectfully.

The present work is just caused by such the necessity. Here, we will
find and describe an invariance group of solutions to the Bogolyubov
equation, as applied to the problem about statistics of random walk
of test particle in thermodynamically equilibrium fluid. Then we will
discuss consequences from exact relations of the group. In
particular, demonstrate that they imply non-Gaussian character of
long-range asymptotic of the random walk. Importantly, in case of gas
this asymptotic, along with above mentioned characteristic time
$\,\sim N\tau\,$, is insensible to gas density and stays strongly
non-Gaussian even in the Boltzmann-Grad limit.

\section{Equations of molecular random walk}

Let a box with volume $\,\Omega\,$ contains $\,N\gg 1\,$ identical
atoms plus one more test particle.  Atoms have mass $\,m\,$,
coordinates $\,{\bf r}_j\,$ and momenta $\,{\bf p}_j\,$
($\,j=1,2...\,N\,$) and interact  with each other via potential
$\,\Phi_a({\bf r}_j-{\bf r}_k)\,$. The test particle has mass
$\,M\,$, coordinate $\,{\bf R}\,$, momentum $\,{\bf P}\,$ and
interacts with atoms via potential $\,\Phi_b({\bf r}_j-{\bf R})\,$.
The potentials are spherically symmetric and short-range with
impenetrable point core.  Because of interactions the test particle
is in chaotic motion, therefore let us name it ``molecular Brownian
particle'' (BP).

We are interested in probability distribution of current position of
BP, $\,{\bf R}(t)\,$, under condition that at initial time moment
$\,t=0\,$ it was placed at certainly known position:\, $\,{\bf
R}(0)={\bf R}_0\,$,\, while personal positions of atoms all the times
are unknown. The simplest statistical ensemble what satisfies this
requirement is determined by the Liouville equation,\, $\,\partial
D_N/\partial t\,=\,[\,H_N\,,D_N\,]\,$,\, for full normalized
distribution function of the system,\, $\,D_N\,$\,, and the initial
condition to it,
\begin{equation}
D_N(\,t=0\,)\,=\,\frac {\delta ({\bf R}-{\bf
R}_0)\,\,e^{-\,H_N/T}}{\int d{\bf R}\int d{\bf P}\int_1...\int_N
\delta ({\bf R}-{\bf R}_0)\,e^{-\,H_N/T}}\,\,\label{din}
\end{equation}
where $\,H_N\,$ is full Hamiltonian of the system (including
interactions with the box walls) and\, $\,\int_k ...=\int\int
...\,\,d{\bf r}_k\,d{\bf p}_k\,$\,. Evidently, such the ensemble
differs from the Gibbs canonic ensemble by the initial BP's
localization only. The latter does not prevent us to introduce the
marginal distribution functions (DF) \,
$\,F_n(t)\,=\,\Omega^{n}\int_{n+1} ...\int_N D_N(t)\,$\, and then go
to the thermodynamical limit ($\,N\rightarrow\infty \,$, $\,\Omega
\rightarrow\infty \,$, $\,\nu=N/\Omega=\,$const\,) in exact analogy
with \cite{bog}. If writing out all the arguments of DF, we have\,
$\,F_n(t)=\,$ $F_n(t,{\bf R}, {\bf r}^{(n)},{\bf P},{\bf
p}^{(n)}|\,{\bf R}_0\,;\nu\,)\,$\,, where\, $\,{\bf r}^{(n)}=\{{\bf
r}_1\,,...\,{\bf r}_n\,\}\,$,\, $\,{\bf p}^{(n)}=\{{\bf
p}_1\,,...\,{\bf p}_n\,\}\,$. As in the Bogolyubov's book, all DF are
not normalized in respect to the atoms' coordinates. Instead, they
must obey the conditions of decoupling of inter-particle correlations
under separation of particles (in essence, that are conditions of
existence of thermodynamical limit \cite{bog,rue}). For our task,
with taking into account complete symmetry of DF in respect to atoms,
these conditions can be written as follows:\,
$\,F_n\,\rightarrow\,F_{n-1}\,G_m({\bf p}_n)\,$\, at \,$\,{\bf
r}_n\rightarrow\infty\,$\,,\, where $\,G_m({\bf p})\,=\,(2\pi
Tm)^{-\,3/2}\exp{(-{\bf p}^2/2Tm)}\,$\, is the Maxwell momentum
distribution of a particle with mass $\,m\,$. The only,
non-principal, difference from \cite{bog} is that numeration of DF
begins from zero, so that $\,F_0(t,{\bf R},{\bf P}|\,{\bf
R}_0\,;\nu\,)\,$ describes the state of BP, and in respect to BP's
coordinate all the DF are normalized. In particular, \, $\,\int\!
F_0\,d{\bf R}\,=\,G_M({\bf P})\,$. The basic Liouville equation
induces the BBGKY equations
\begin{equation}
\frac {\partial F_n}{\partial t}=[\,H_{n}\,,F_n\,]\,+\,\nu\, \frac
{\partial }{\partial {\bf P}}\int_{n+1}\!\! \Phi^{\,\prime}_b({\bf
R}-{\bf r}_{n+1})\,F_{n+1}\,+\,\nu \sum_{j\,=1}^n\,\frac {\partial
}{\partial {\bf p}_j}\int_{n+1}\!\!\Phi^{\,\prime}_a({\bf r}_j-{\bf
r}_{n+1}) \,F_{n+1}\,\label{fn}
\end{equation}
($\,n\,=\,0,\,1,\,\dots\,$)\, with initial conditions
\begin{equation}
\begin{array}{l}
F_n(t=0)\,=\, \delta({\bf R}-{\bf R}_0)\,F_n^{(eq)}({\bf
r}^{(n)}\,|{\bf R};\nu)\,G_M({\bf P})\prod_{j\,=1}^n G_m({\bf
p}_j)\,\,\,,\label{ic}
\end{array}
\end{equation}
where\, $\,H_{n}\,$ is Hamiltonian of subsystem ``$\,n\,$ atoms plus
BP\,'', $\,\Phi^{\,\prime}_{a,\,b}({\bf r})=\nabla\Phi_{a,\,b}({\bf
r})\,$\,,\, and\, $\,F_n^{(eq)}({\bf r}^{(n)}\,|{\bf R};\nu)\,$ are
usual thermodynamically equilibrium DF for $\,n\,$ atoms in presence
of BP occupying point $\,{\bf R}\,$.

In principle, all that will do for finding $\,F_0(t,{\bf R},{\bf
P}|\,{\bf R}_0\,;\nu\,)\,$ and thus probability distribution of BP's
path,\, $\,\Delta{\bf R}(t)={\bf R}(t)-{\bf R}_0\,$,\, without any
additional assumptions. At that, since our problem is expressed by
the recurrent relations, it can be lighten, as usually, by use of
generating functions. Such the approach to the BBGKY hierarchy was
formulated and sampled already by Bogolyubov \cite{bog}. Here it will
help us to visualize some properties of the hierarchy what are hardly
seen directly from it. Following Bogolyubov, let us combine all DF
into generating functional (GF)
\begin{equation} \mathcal{F}\{t,{\bf
R},{\bf P},\psi\,|{\bf R}_0;\nu\}\,=\,F_0\,+\sum_{n\,=1}^{\infty }
\frac {\nu^n}{n!}\int_1 ...\int_n F_n \,\prod_{j\,=1}^n \psi({\bf
r}_j,{\bf p}_j)\,\,\label{gf}
\end{equation}
and equations (\ref{fn}) into corresponding
``generating equation'' for it:
\begin{eqnarray}
\frac {\partial \mathcal{F}}{\partial t}\,+\,\frac {\bf
P}{M}\cdot\frac {\partial \mathcal{F}}{\partial {\bf
R}}\,=\,\mathcal{\widehat{L}}\left(\psi,\frac {\delta
}{\delta\psi}\right)\,\mathcal{F}\,\,\,,\label{fe}
\end{eqnarray}
where $\,\mathcal{\widehat{L}}\,$ is operator composed of usual and
variational derivatives,
\begin{eqnarray}
\mathcal{\widehat{L}}\left(\psi,\frac {\delta
}{\delta\psi}\right)\,=\,-\int_1 \psi(x_1)\,\, \frac {{\bf
p}_1}{m}\cdot\frac {\partial }{\partial {\bf r}_1}\,\frac {\delta
}{\delta \psi(x_1)}\,\,+\,\;\;\;\;\;\;\;\;\;\;\;\;\;
\;\;\;\;\;\;\;\;\;\;\;\;\;\;\; \;\;\;\;\;\;\;\;\;\;
\;\;\;\;\;\;\;\;\;\;\; \label{l}\\
+\,\,\int_1\,\, [\,1+\psi(x_1)\,]\left[\,\Phi_b({\bf R}-{\bf
r}_{1})\,,\frac {\delta }{\delta
\psi(x_1)}\,\right]\,+\,\;\;\;\;\;\;\;\;\;\;\;\;\;\;\;\;\;
\;\;\;\;\;\;\;\;\;\;\;\;\;\nonumber\\
+\,\frac 12 \int_1\int_2\,\, [\,1+\psi(x_1)\,]\,[\,1+\psi(x_2)\,]
\left[\,\Phi_a({\bf r}_1-{\bf r}_2)\,,\frac {\delta^{\,2} }{\delta
\psi(x_1)\,\delta \psi(x_2)}\,\right]\,\,\,,\nonumber
\end{eqnarray}
with\,  $\,x_j\,=\,\{{\bf r}_j,{\bf p}_j\}\,$\,. This is direct
analogue of equation  (7.9) from \cite{bog}. To make this evident,
notice that $\,\psi(x)=u(x)/\nu\,$, where $\,u(x)\,$ is functional
argument used in  \cite{bog}, and factor
$\,[1+\psi(x_1)]\,[1+\psi(x_2)]\,$ can be replaced by
$\,[\,\psi(x_1)+\psi(x_2)+\psi(x_1)\psi(x_2)\,]\,$ due to the
identity\, $\,\int_1\int_2 \,[\,\Phi_a({\bf r}_1-{\bf r}_2)\,,
...\,]=0\,$.\,

Let us formulate the initial condition to equation (\ref{fe}). For
that, it is convenient to introduce the generating functional of
equilibrium DF in the configuration space,
\begin{eqnarray}
\mathcal{F}^{(eq)}\{\phi |\,{\bf R};\nu\}\,= \,1\,
+\sum_{n\,=1}^{\infty }\frac {\nu^n}{n!}\int\! ...\!\int
F_n^{(eq)}({\bf r}^{(n)}|\,{\bf R};\nu)\, \prod_{j\,=1}^n \phi({\bf
r}_j)\,d{\bf r}_j\,\,\,,\label{eqf}
\end{eqnarray}
where $\,\phi({\bf r})\,$ is corresponding functional variable.
Besides, introduce linear mapping $\,\phi\{\psi\}\,$ of functions
$\,\psi({\bf r},{\bf p})\,$ to functions $\,\phi({\bf r})\,$\, as
follows:
\begin{eqnarray}
\phi\{\psi\}({\bf r})\,=\,\int \psi({\bf r},{\bf p})\,G_m({\bf
p})\,d{\bf p}\,\label{phi}
\end{eqnarray}
Then the initial conditions (\ref{ic}) do intend
\begin{eqnarray}
\mathcal{F}\{0,\,{\bf R},{\bf P},\psi\,|\,{\bf R}_0;\nu\}\, =
\,\delta({\bf R}-{\bf R}_0)\,G_M({\bf P})\,
\mathcal{F}^{(eq)}\{\phi\{\psi\} |\,{\bf R};\nu\}\,\,\label{icf}
\end{eqnarray}
It is easy to guess or verify that expression\, $\,G_M({\bf
P})\,\mathcal{F}^{(eq)}\{\phi\{\psi\} |\,{\bf R};\nu\}\,$\,
represents the stationary solution of equation (\ref{fe}), so that
\[
\left[-({\bf P}/M)\cdot\partial /\partial {\bf
R}+\mathcal{\widehat{L}}\,\right ]\,G_M({\bf P})\,
\mathcal{F}^{(eq)}\{\phi\{\psi\} |\,{\bf R};\nu\}\,=\,0\,
\]
This equality yields
\begin{equation} \left[\frac {\partial }{\partial
{\bf r}}\,+\frac {\Phi_b^{\,\prime}({\bf r}-{\bf R})}{T}\right]\frac
{\delta \mathcal{F}^{(eq)}}{\delta \phi({\bf r})}\,=\,\frac 1T \int
[\,1+\phi({\bf r}^{\prime})\,]\,\Phi_a^{\,\prime}({\bf
r}^{\prime}-{\bf r})\, \frac {\delta^2 \mathcal{F}^{(eq)}}{\delta
\phi({\bf r})\,\delta \phi({\bf r}^{\prime})}\,d{\bf
r}^{\prime}\,\,\,,\label{ter}
\end{equation}
where, similarly to (\ref{eqf}),\, $\,\phi({\bf r})\,$ appears as
independent functional variable in the configuration space. The
latter equation determines all the equilibrium DF and is analogue of
equation (2.14) from \cite{bog}.

Unfortunately, to the best of my knowledge, non-stationary solutions
to equations like (\ref{fe}) above  or (7.9) in \cite{bog} never were
investigated by Bogolyubov or others. However, the experience of work
with the BBGKY equations prompts (see e.g. \cite{sil}) the
desirability of a change of variables, i.e. transition from DF to
another functions which help to concentrate on inter-particle
correlations and statistical dependencies. To make a suitable choice
of such functions, let us discuss hypothetical equalities
\[
\begin{array}{l}
F_n(t,{\bf R}, {\bf r}^{(n)},{\bf P},{\bf p}^{(n)}|\,{\bf
R}_0;\nu)\,\stackrel{?}{=}\,F_0(t,{\bf R},{\bf P}|\,{\bf R}_0;\nu)
\,\,F_n^{(eq)}({\bf r}^{(n)}|\,{\bf R};\nu)\prod_{j\,=1}^n G_m({\bf
p}_j)\,\,
\end{array}
\]
or, equivalently, \,\, $\,\mathcal{F}\{t,{\bf R},{\bf
P},\,\psi\,|\,{\bf R}_0;\nu\}\,\stackrel{?}{=}\,F_0(t,{\bf R},{\bf
P}|\,{\bf R}_0;\nu)\,\mathcal{F}^{(eq)}\{\phi\{\psi\}\,|\,{\bf
R};\nu\}\,$. They state, evidently, that the conditional DF
(conditional probability distributions) of atoms, i.e.\,
$\,F_n/F_0\,$\,, do not depend on the past BP's displacement  $\,{\bf
R}-{\bf R}_0\,$. This assumption seems natural as concerns
thermodynamically equilibrium random walk when all possible positions
of BP are equivalent. Nevertheless, this is certainly wrong
assumption, since it is incompatible with the equations (\ref{fn}).
Indeed, the substitution of $\,F_1\,$ in the first equation (for
$\,F_0\,$) by\, $\,F_0(t,{\bf R},{\bf P}|\,{\bf R}_0;\nu)
\,F_1^{(eq)}({\bf r}_1|\,{\bf R};\nu) G_m({\bf p}_1)\,$ turns the
integral of BP's interaction with atoms into zero:\, $\,\int_1
\Phi^{\,\prime}_b({\bf R}-{\bf r}_1)\,F_{1}=0\,$ (at least because
$\,F_1^{(eq)}({\bf r}_1|{\bf R};\nu)\,$ is even function of the
difference $\,{\bf r}_1-{\bf R}\,$ while\, $\,\Phi^{\,\prime}_b({\bf
R}-{\bf r}_1)\,$ is its odd function). As the result, the first of
equations (\ref{fn}) reduces to the equation of free BP's flight,
$\,\partial F_0/\partial t =[H_{0}\,,F_0]\,$, as if BP does not
interact with atoms at all.

The aforesaid implies that, first, the current state of the medium
(system of atoms) is statistically dependent on the summary
displacement of BP $\,\Delta{\bf R}(t)={\bf R}(t)-{\bf R}_0\,$ during
all time of its observation. In other words, some essential and
quantitatively significant correlations between the medium and the
history of BP's wandering are in existence. We will name them
``historical correlations'', in order to ideally separate them from
the equilibrium correlations, as described by $\,F_n^{(eq)}({\bf
r}^{(n)}|\,{\bf R};\nu)\,$, between current coordinates of BP and
atoms. Second, to perform an adequate formal separation of the two
sorts of BP-atoms correlations, we can describe the historical
correlations with functions\, $\,V_n=V_n(t,{\bf R}, {\bf
r}^{(n)},{\bf P},{\bf p}^{(n)}|\,{\bf R}_0;\nu)\,$\, defined by the
generating relations as follow:
\begin{eqnarray}
\mathcal{F}\{t,{\bf R},{\bf P},\,\psi\,|\,{\bf R}_0;\nu\}\,=
\,\mathcal{V}\{t,{\bf R},{\bf P},\,\psi\,|\,{\bf
R}_0;\nu\}\,\,\mathcal{F}^{(eq)}\{\phi\{\psi\}\,|\,{\bf
R};\nu\}\,\,\,,\label{vf}\\
\mathcal{V}\{t,{\bf R},{\bf P},\psi\,|\,{\bf
R}_0;\nu\}\,=\,V_0\,+\sum_{n\,=1}^{\infty } \frac {\nu^n}{n!}\int_1
...\int_n V_n \,\prod_{j\,=1}^n \psi({\bf r}_j,{\bf
p}_j)\,\,\nonumber
\end{eqnarray}
In particular,\, $\,V_0(t,{\bf R},{\bf P}| \,{\bf
R}_0;\nu)\,=\,F_0(t,{\bf R},{\bf P}|\, {\bf R}_0;\nu)\,$\, and
\begin{equation}
\begin{array}{l}
F_1(t,{\bf R},{\bf r}_1,{\bf P},{\bf p}_1|\,{\bf
R}_0;\nu)\,=\,\\
\,\,\,\,\,\,\,\,\,=\,F_0(t,{\bf R},{\bf P}|\,{\bf
R}_0;\nu)\,F_1^{(eq)}({\bf r}_1|{\bf R};\nu)\,G_m({\bf p}_1)\,+\,
V_1(t,{\bf R},{\bf r}_1,{\bf P},{\bf p}_1|\,{\bf R}_0;\nu)\,
\label{cf1}
\end{array}
\end{equation}
It is clear from this definition that from the viewpoint of the
probability theory the functions \, $\, V_n\,$ (при $\,n>0\,$)
represent a kind of cumulant functions (cumulants, or
semi-invariants). Therefore we will name them ``cumulant functions''.
Notice that in physical literature (see e.g. \cite{sil}) similar
objects frequently are called ``correlation functions''.

In terms of the cumulant functions (CF) initial conditions (\ref{ic})
and (\ref{icf}) become strongly simplified:\,
\begin{eqnarray}
V_0(0\,,{\bf R},{\bf P}|\,{\bf R}_0;\nu)\,\,=\,\delta({\bf R}-{\bf
R}_0)
\,G_M({\bf P})\,\,\,,\nonumber\\
V_{n}(0\,,{\bf R}, {\bf r}^{(n)},{\bf P},{\bf p}^{(n)}|\,{\bf
R}_0;\nu)\,\,=\,0\,\,\,\,\,\,(n>0)\,\,\,,\nonumber\\
\mathcal{V}\{0,\,{\bf R},{\bf P},\,\psi|\,{\bf
R}_0;\nu\}\,=\,\delta({\bf R}-{\bf R}_0)\,G_M({\bf P})\, \label{icv}
\end{eqnarray}
The conditions of the decoupling of correlations at infinity also
become simpler:
\begin{equation}
\begin{array}{l}
V_n(t\,,{\bf R}, {\bf r}^{(n)},{\bf P},{\bf p}^{(n)}|\,{\bf
R}_0;\nu)\,\rightarrow\,0\,\,\,\,\,\,\,\texttt{at}\,\,\,\,\,\,{\bf
r}_k\rightarrow \infty\,\,\label{bcv}
\end{array}
\end{equation}
($\,1\leq k\leq n\,$), that is CF tend to zero if at least one
$\,n\,$ atoms moves away from BP to infinity. In opposite, the
generating equation (\ref{fe}) in terms of CF becomes more
complicated. Inserting (\ref{vf}) into (\ref{fe}) one can obtain
\begin{equation}
\frac {\partial \mathcal{V}}{\partial t}\,+\,\frac {\bf
P}{M}\cdot\frac {\partial \mathcal{V}}{\partial {\bf
R}}\,=\,\widehat{\mathcal{L}}\left(\psi,\frac {\delta
}{\delta\psi}\right)\,\mathcal{V}\,+
\,\widehat{\mathcal{L}}^{\,\,\prime}\left(\nu,\psi,\frac {\delta
}{\delta\psi}\right)\,\mathcal{V}\,\,\,,\,\label{fev}
\end{equation}
where the new operator appears,
\begin{eqnarray}
\widehat{\mathcal{L}}^{\,\,\prime}\left(\nu,\psi,\frac {\delta
}{\delta\psi}\right)\,=\,\left\{\int [\,1\,+\phi({\bf
r})\,]\,\,\Phi_b^{\,\prime}({\bf R}-{\bf
r})\,\,\nu\,\mathcal{C}\{{\bf r},\phi\,|\,{\bf R};\,\nu\}\,\,d{\bf
r}\right\}\left(\frac {{\bf P}}{MT}+\frac {\partial
}{\partial {\bf P}}\right )\,+\,\,\,\,\nonumber\\
+\,\int_1\int_2\,\,
[\,1+\psi(x_1)\,]\,[\,1+\psi(x_2)\,]\left[\,\Phi_a({\bf r}_1-{\bf
r}_2)\,,\,\nu\,\mathcal{C}\{{\bf r}_2,\phi\,|\,{\bf
R};\,\nu\}\,G_m({\bf p}_2)\,\frac {\delta }{\delta \psi(x_1)}
\,\right]\nonumber\,\,
\end{eqnarray}
Here\, $\,\phi({\bf r})\,$ and $\,\phi\,$ are mentioned as linear
functionals of $\,\psi({\bf r},{\bf p})\,$ in the sense of the
mapping\, $\,\phi\{\psi\}({\bf r})\,$\, defined by (\ref{phi}), and
besides one more functional is introduced,
\begin{equation}
\mathcal{C}\{{\bf r},\phi\,|\,{\bf R};\,\nu\}\,=\,\frac {\delta \ln
\mathcal{F}^{(eq)}\{\phi |\,{\bf R};\,\nu\}}{\nu\,\delta \phi({\bf
r})}\,\label{c} \,
\end{equation}
Correspondingly, the BBGKY equations become more complicated.
Therefore, here we write out them only for extreme but principally
interest case of ``BP in ideal gas'' (when atoms do not interact with
each other, i.e. $\,\Phi_a({\bf r})=0\,$). In this case,
\begin{eqnarray}
\frac {\partial V_0}{\partial t}\,=\,-\frac {\bf P}{M}\cdot\frac
{\partial V_0}{\partial {\bf R}}\,+\,\nu\, \frac {\partial }{\partial
{\bf P}}\int_{1} \Phi^{\prime}_b({{\bf R}-\bf
r}_{1})\,V_{1}\,\,\,,\nonumber
\end{eqnarray}
\begin{eqnarray}
\frac {\partial V_{n}}{\partial t}\,=\,[\,H_n\,,V_n\,]\,+\,\nu\,
\frac {\partial }{\partial {\bf P}}\int_{n+1} \Phi^{\prime}_b({{\bf
R}-\bf r}_{n+1})\,V_{n+1}\,\,+\,\,\,\,\,\,\,\,\,\,\,\,
\,\,\,\,\,\,\label{vn}
\end{eqnarray}
\begin{eqnarray}
\,\,\,\,\,\,\,\,\,\,\,\,\,\,\,\,\,+\,\,T\,\sum_{j\,=1}^{n}\,
\mathcal{P}(j,n)\,\,G_m({\bf p}_n)\,E^{\,\prime}({\bf r}_n-{\bf
R})\left(\frac {{\bf P}}{MT}+\frac {\partial }{\partial {\bf
P}}\right ) V_{n-\,1}\,\,\,\,\,\,\,\,\,(n>0)\,\,\,,\nonumber
\end{eqnarray}
where\, $\,H_n ={\bf P}^2/2M + \sum_{j\,=1}^n\,[\,{\bf p}_j^2/2m
+\Phi_b({{\bf R}-\bf r}_j)\,]\,$\,,\,  $\,E({\bf
r})=\exp{[\,-\,\Phi({\bf r})/T\,]}\,$,\, $\,E^{\,\prime}({\bf
r})=\nabla E({\bf r})= -\,\Phi^{\,\prime }({\bf r})E({\bf r})/T\,$\,,
\,and\, $\,\mathcal{P}(j,n)\,$\, denotes operation of transposition
of arguments\, $\,x_j\,=\,\{{\bf r}_j,{\bf p}_j\}\,$\, and
$\,x_n\,=\,\{{\bf r}_n,{\bf p}_n\}\,$.\, Thus, the natively
bidiagonal BBGKY hierarchy if being represented via CF becomes al
least tridiagonal.

\section{Invariance group of the equilibrium generating functional}
Further, let us consider the functional (\ref{c}), which is
interesting already because it enters the generating evolution
equation (\ref{fev}), and expand it into a series:
\begin{equation}
\mathcal{C}\{{\bf r},\phi\,|\,{\bf R};\,\nu\}=F_1^{(eq)}({\bf
r}|\,{\bf R};\nu)+\sum_{n\,=\,1}^{\infty }\frac {\nu^n}{n!}\int\!\!
...\!\!\int C_{n+1}({\bf r},{\bf r}_1...\,{\bf r}_n|\,{\bf R};\nu)
\prod_{j\,=1}^n \phi({\bf r}_j)\,d{\bf r}_j\,\label{c1}
\end{equation}
According to (\ref{eqf}), this expansion just defines the functions\,
$\,C_{n}({\bf r}_1...\,{\bf r}_n|\,{\bf R};\nu)\,$ ($\,n>1\,$). In
particular,
\[
\begin{array}{l}
C_2({\bf r},{\bf r}_1|\,{\bf R};\nu)\,= \,F_2^{(eq)}({\bf r},{\bf
r}_1|{\bf R};\nu)-F_1^{(eq)}({\bf r}|{\bf
R};\nu)\,F_1^{(eq)}({\bf r}_1|{\bf R};\nu)\,\,\,,\\
C_3({\bf r},{\bf r}_1,{\bf r}_2)\,=\,F^{(eq)}_3({\bf r},{\bf
r}_1,{\bf r}_2)\,+\,2\,F^{(eq)}_1({\bf r})\,F^{(eq)}_1({\bf
r}_1)\,F^{(eq)}_1({\bf r}_2)\,-\\
\,\,\,\,\,\,\,\,\,\,\,\,\,\,\,\, \,\,\,\,\,\,\,-\,F^{(eq)}_1({\bf
r})\,F^{(eq)}_2({\bf r}_1,{\bf r}_2)\,-\,F^{(eq)}_1({\bf
r}_1)\,F^{(eq)}_2({\bf r},{\bf r}_2)\,-\,F^{(eq)}_1({\bf
r}_2)\,F^{(eq)}_2({\bf r},{\bf r}_1)\,\,\,,
\end{array}
\]
(for brevity in the second expression arguments $\,{\bf R}\,$ and
$\,\nu\,$ are omitted). Clearly, from the viewpoint of the atoms'
coordinates, the functions $\,C_{n}\,$ relate to the equilibrium DF
$\,F^{(eq)}_n\,$\, like cumulants of a random field relate to its
statistical moments.

The conditions of vanishing of inter-particle correlations at
infinity mean that all these cumulants tend to zero if at least one
atom is moved away from the others:
\begin{equation}
\begin{array}{l}
C_{n}({\bf r}_1\,...\,{\bf r}_n|\,{\bf
R};\nu)\,\rightarrow\,0\,\,\,\,\,\,\,\,\,\,\,\texttt{at}
 \,\,\,\,\,\,\,\,\,{\bf r}_j\rightarrow\infty \,\label{cbc}
\end{array}
\end{equation}
In case of mowing away BP from atoms we have:
\begin{equation}
\begin{array}{l}
F_1^{(eq)}({\bf r}|\,{\bf R};\nu) \,\rightarrow\,1\,
\,\,\,\,\,\,\,\,\,\,\texttt{at}
 \,\,\,\,\,\,\,\,\,{\bf R}\rightarrow\infty \,\,\,,\\
C_{n}({\bf r}_1\,...\,{\bf r}_n|\,{\bf R};\nu)\,\rightarrow\,
C_{n}({\bf r}_1\,...\,{\bf r}_n;\,\nu)
\,\,\,\,\,\,\,\,\,\,\texttt{at}
 \,\,\,\,\,\,\,\,\,{\bf R}\rightarrow\infty \,\,\,,\label{cbc1}
\end{array}
\end{equation}
where\, $\,C_{n}({\bf r}_1\,...\,{\bf r}_n;\,\nu)\,$\, are cumulant
functions of equilibrium media taken in absence of BP. Moreover,
under sufficiently short-range interaction potentials, all the limit
transitions in (\ref{cbc})-(\ref{cbc1}) are absolutely integrable,
which will be assumed below.

Next, consider equation  (\ref{ter}) rewriting it in the form
\[
\left[\frac {\partial }{\partial {\bf r}}+\frac
{\Phi_b^{\,\prime}({\bf r}-{\bf R})}{T}\right]\mathcal{C}\{{\bf
r},\phi\,|{\bf R};\,\nu\}\,=\,\frac 1T \int [\,1+\phi({\bf
r}^{\prime})\,]\, \Phi_a^{\,\prime}({\bf r}^{\prime}-{\bf r})\,\frac
{\delta \mathcal{C}\{{\bf r}^{\,\prime},\phi\,|{\bf
R};\,\nu\}}{\delta \phi({\bf r})}\,\,d{\bf r}^{\prime}\,+\nonumber
\]
\begin{eqnarray}
+\,\,\mathcal{C}\{{\bf r},\phi\,|{\bf R};\,\nu\}\,\,\frac
{\nu}{T}\int [\,1+\phi({\bf r}^{\prime})\,]\,\Phi_a^{\,\prime}({\bf
r}^{\prime}-{\bf r})\,\mathcal{C}\{{\bf r}^{\prime},\phi\,|{\bf
R};\,\nu\}\,d{\bf r}^{\,\prime}\,\,\,,\,\label{ter1}
\end{eqnarray}
and with its help let us ascertain (probably, for the first time)
some important properties of the functional\, $\,\mathcal{C}\{{\bf
r},\phi\,|\,{\bf R};\,\nu\}\,$. Notice, first, that integrability of
the asymptotic (\ref{cbc}) makes it possible to extend this
functional to bounded functions $\,\phi({\bf r})\,$ which do not tend
to zero at infinity, in particular, to constants. This fact allows to
introduce such the objects as follow:
\begin{equation}
C(\sigma ,\nu)=\lim_{\begin{array}{c}
                      \phi({\bf r})\rightarrow\sigma  \\
                       {\bf R}\,\rightarrow\infty
                     \end{array}
} \mathcal{C}\{{\bf r},\phi|\,{\bf R};\nu\}=1+\sum_{n\,=\,1}^{\infty
}\frac {\nu^n\sigma^n}{n!}\int_1\! ...\!\int_n C_{n+1}({\bf r},{\bf
r}_1\,...\,{\bf r}_n;\nu)\,\,\,,\label{ccc}
\end{equation}
\begin{equation}
\mathcal{C}_{\sigma}\{{\bf r},\phi\,|{\bf R};\nu\}\,=\,\frac
{\mathcal{C}\{{\bf r},\sigma +\phi\,|\,{\bf R};\,\nu\}}{C(\sigma
,\nu)}\,\,\,,\label{nc}
\end{equation}
where\, $\,\sigma=\,$const\,,\, $\,\int_n...\,=\,\int...\,\,d{\bf
r}_n\,$\,, and integrals in (\ref{ccc}) are practically independent
on $\,\,{\bf r}\,$. Second, after the change\, $\,\phi({\bf
r})\rightarrow\sigma +\phi({\bf r})\,$ the equation (\ref{ter1}) can
be transformed, through elementary algebraic manipulations, into
equation for $\,\mathcal{C}_{\sigma}\{{\bf r},\phi\,|{\bf
R};\nu\}\,$\,:
\[
\left[\frac {\partial }{\partial {\bf r}}+\frac
{\Phi_b^{\,\prime}({\bf r}-{\bf
R})}{T}\right]\mathcal{C}_{\sigma}\{{\bf r},\phi\,|{\bf
R};\nu\}=\frac 1T\! \int \!\!\left[1+\frac {\phi({\bf
r}^{\prime})}{1+\sigma}\right]\Phi_a^{\,\prime}({\bf r}^{\prime}-{\bf
r})\,\frac {\delta\, \mathcal{C}_{\sigma}\{{\bf
r}^{\,\prime},\phi\,|{\bf R};\nu\}}{\delta\,[\, \phi({\bf
r})/(1+\sigma )\,]}\,\,d{\bf r}^{\prime}\,+
\]
\begin{equation}
+\,\mathcal{C}_{\sigma}\{{\bf r},\phi\,|{\bf R};\nu\}\,\,\frac
{\nu\,C(\sigma ,\nu)\,(1+\sigma)}{T}\int\! \left[1+\frac {\phi({\bf
r}^{\prime})}{1+\sigma}\right]\Phi_a^{\,\prime}({\bf r}^{\prime}-{\bf
r})\,\,\mathcal{C}_{\sigma}\{{\bf r}^{\,\prime},\phi\,|{\bf
R};\nu\}\,\,d{\bf r}^{\,\prime}\,\,\label{ter2}
\end{equation}
It is easy to see that its only difference from (\ref{ter1}) is the
scale transformation of the functional argument,\, $\,\phi({\bf
r})\rightarrow \phi({\bf r})/(1+\sigma )\,$\,, and besides
transformation of the density argument\, $\,\nu\,$\, into
\begin{equation}
\begin{array}{l}
\upsilon (\sigma ,\nu)\,=\,\nu\,C(\sigma
,\nu)\,(1+\sigma)\,\,\label{nd}
\end{array}
\end{equation}
Third, formal solution of equation (\ref{ter1}) in the form of the
series (\ref{c1}) is unambiguously determined by the conditions of
decoupling of correlations (\ref{cbc})-(\ref{cbc1}). Fourth, by the
definition (\ref{ccc})-(\ref{nc}) of functional
$\,\mathcal{C}_{\sigma}\{{\bf r},\phi\,|{\bf R};\nu\}\,$, the
coefficients in its similar series expansion in terms of $\,\phi({\bf
r})\,$ satisfy the same conditions. To be exact, from
(\ref{cbc})-(\ref{cbc1}) it follows that
\begin{eqnarray}
C_{n}({\bf r}_1\,...\,{\bf r}_{n}|\,{\bf
R};\nu)+\sum_{k\,=1}^{\infty}\frac {\nu^k\sigma^k}{k!} \int_{n+1}\!
...\!\int_{n+k} C_{n+k}({\bf r}_1\,...\,{\bf r}_{n+k}|\,{\bf
R};\nu)\,\rightarrow\,0\, \,\,\,\,\,\texttt{at}\,\,\,\,\,{\bf
r}_j\rightarrow\infty\,\,,\nonumber\\
\frac {1}{C(\sigma ,\nu)}\left[F_1^{(eq)}({\bf r}|\,{\bf
R};\nu)+\sum_{k\,=1}^{\infty}\frac {\nu^k\sigma^k}{k!} \int_{1}\!
...\!\int_{k} C_{k+1}({\bf r},{\bf r}_1\,...\,{\bf r}_{k}|\,{\bf
R};\nu)\right]\,\rightarrow\,1\, \,\,\,\,\texttt{at}\,\,\,\,{\bf
R}\rightarrow\infty\,\,,\nonumber
\end{eqnarray}
where\, $\,1\leq j\leq n\,$. At least, if we interpret the
transitions (\ref{cbc})-(\ref{cbc1}) in the sense of absolute
integrability and speak about not too large density values (useful
information in this respect can be found in \cite{rue}).

Summarizing all the aforesaid, we can conclude that the solution to
equation (\ref{ter2}) is nothing but\,\,
$\,\mathcal{C}_{\sigma}\{{\bf r},\phi\,|{\bf R};\nu \}\,=
\,\mathcal{C}\{{\bf r},\phi/(1+\sigma )\,|\,{\bf R};\,\upsilon
(\sigma ,\nu) \}\,$\,.\, Combining this result with (\ref{nc}) and
(\ref{nd}), we see that for arbitrary, in definite sense (see below),
constant $\,\sigma \,$ and bounded function $\,\phi=\phi({\bf r})\,$
the equality
\begin{equation}
\nu\,\mathcal{C}\{{\bf r},\sigma +\phi\,|\,{\bf R};\,\nu\}\,=\,\frac
{\upsilon (\sigma ,\nu)}{1+\sigma}\,\,\mathcal{C}\left\{{\bf r},\frac
{\phi}{1+\sigma}\,|\,{\bf R};\,\upsilon (\sigma
,\nu)\right\}\,\,\,\label{eqid}
\end{equation}
takes place. It can be rewritten also as
\begin{equation}
\widehat{\mathcal{T}}(\sigma)\,\mathcal{C}\{{\bf r},\phi\,|\,{\bf
R};\,\nu\}\,\equiv\,\frac {\upsilon (\sigma ,\nu)}{(1+\sigma )\,\nu
}\,\,\mathcal{C}\left\{{\bf r},\frac {1+\phi}{1+\sigma}-1\,|\,{\bf
R};\,\upsilon(\sigma ,\nu)\right\}\,= \,\mathcal{C}\{{\bf
r},\phi\,|\,{\bf R};\,\nu\}\,\label{gr}
\end{equation}
The left equality here defines the one-parametric family of such
transformations of the functional $\,\,\mathcal{C}\{{\bf
r},\phi\,|\,{\bf R};\,\nu\}\,$ which, according to the right-hand
equality, conserve its value. One can easy verify that this family is
the Abelian group with composition rules
\begin{eqnarray}
\widehat{\mathcal{T}}(\sigma_2)\,\widehat{\mathcal{T}}(\sigma_1)\,=
\,\widehat{\mathcal{T}}(\sigma_1 +\sigma_2 +\sigma_1\sigma_2)
\,\,\,,\nonumber\\ \upsilon (\sigma_2\,,\,\upsilon (\sigma_1\,
,\nu))\,=\,\upsilon (\sigma_1 +\sigma_2 +\sigma_1\sigma_2
\,,\nu)\,\,\label{gr1}
\end{eqnarray}
and with restrictions\, $\,\sigma >-1\,$\,,\, $\,\phi({\bf r})
>-1\,$\,. The latter are clear in view of the fact that functional\,
$\,\,\nu\,[1+\phi({\bf r})]\,\mathcal{C}\{{\bf r},\phi\,|\,{\bf
R};\,\nu\}\,$\, represents mean concentration of atoms in presence of
an external potential $\,U({\bf r})\,$ related to $\,\phi({\bf r})\,$
via\, $\,\phi({\bf r})=\exp{[\,-\,U({\bf r})/T\,]}-1\,$\, (see e.g.
\cite{mpa3}). The substitution $\,\sigma =\exp{(a)}-1\,$ yields\,
$\,\widehat{\mathcal{T}}(a_2)\,\widehat{\mathcal{T}}(a_1)=
\widehat{\mathcal{T}}(a_1 +a_2)\,$\, and eliminates the restrictions.

The infinitesimal equivalent of the equalities (\ref{eqid}) or
(\ref{gr}) looks most convenient when expressed via particular CF\,:
\begin{eqnarray}
\left\{\varkappa(\nu)+[\,1+\varkappa(\nu)\,] \,\nu\,\frac
{\partial}{\partial \nu}\right\}F_1^{(eq)}({\bf r}|\,{\bf
R};\nu)\,=\,\nu\!\int\! C_2({\bf r},{\bf r}^{\,\prime}|{\bf
R};\nu)\,d{\bf r}^{\,\prime}\,\,\,,
\,\,\,\,\,\,\,\,\,\,\,\,\,\label{eqinf}\\
\left\{n\varkappa(\nu)+[\,1+\varkappa(\nu)\,] \,\nu\,\frac
{\partial}{\partial \nu}\right\} C_{n}({\bf r}_1\,...\,{\bf
r}_n|\,{\bf R};\nu)=\nu\!\int\! C_{n\,+1}({\bf r}_1...\,{\bf
r}_{n},{\bf r}^{\,\prime}|{\bf R};\nu)\,d{\bf r}^{\,\prime}
\,\,\,,\nonumber\\
\varkappa(\nu)\,\equiv \,\left[\frac {\partial C(\sigma
,\nu)}{\partial \sigma}\right]_{\sigma =0}\,=\,\nu\int C_2({\bf
r},0;\,\nu)\,d{\bf r}\,\,\,\,\,\,\,\,\,\,\,\,\,\,
\,\,\,\,\,\,\,\,\,\, \,\,\,\,\,\,\,\,\,\, \,\,\,\,\, \nonumber
\end{eqnarray}
The function\, $\,\varkappa(\nu)\,$\, defined just now is known (see
e.g.\cite{ll1}) to be directly concerning the state equation of the
system:\, $\,1+\varkappa(\nu)\,=\,T\,(\partial \nu/\partial
\mathcal{P})_T\,$\,, where\, $\,\mathcal{P}\,$ denotes the pressure.
Notice that in the framework of the grand canonical ensemble very
similar relations can be easy derived by differentiation of DF in
respect to the activity.

\section{Invariance group of the generating functional \\
of historical correlations} We have approached to the substantiation
of main results of the present work. Let us show that solution of the
evolution equation (\ref{fev}) possesses invariance properties
similar to (\ref{gr}). Since the initial condition to (\ref{fev})
(see (\ref{icv})) does not depend on the variables $\,\psi =\psi({\bf
r},{\bf p})\,$ and $\,\nu\,$\, at all, the solution to (\ref{fev}) is
completely determined by the structure of operators
$\,\widehat{\mathcal{L}}\,$ and
$\,\widehat{\mathcal{L}}^{\,\,\prime}\,$ and the conditions at
infinity (\ref{bcv}). The latter allow to extend the functional
$\,\mathcal{V}\{t,{\bf R},{\bf P},\psi\,|{\bf R}_0;\nu\}\,$ (like the
functional $\,\mathcal{C}\,$ before) to arguments  $\,\sigma
+\psi({\bf r},{\bf p})\,$, with $\,\sigma =\,$\,const\,, in place of
$\,\psi({\bf r},{\bf p})\,$. The fact that the limit in (\ref{bcv})
is achieved rapidly enough to indeed ensure this extension will
become clear afterwards, from the relations (\ref{inf}) which will be
obtained in this section. Next, thanks to (\ref{bcv}) and besides to
the definition (\ref{l}) of operator $\,\widehat{\mathcal{L}}\,$\,,
the variable $\,\psi(x_1)\,$ in the first term of the expression\,
$\,\widehat{\mathcal{L}}\,\mathcal{V}\,$\, can be shifted by
arbitrary constant:
\[
\int_1 \psi(x_1)\,\, \frac {{\bf p}_1}{m}\cdot\frac {\partial
}{\partial {\bf r}_1}\,\frac {\delta \mathcal{V}}{\delta
\psi(x_1)}\,\,=\,\int_1 \,[\,a +\psi(x_1)\,]\,\, \frac {{\bf
p}_1}{m}\cdot\frac {\partial }{\partial {\bf r}_1}\,\frac {\delta
\mathcal{V}}{\delta \psi(x_1)}\,\,\,,
\]
where\, $\,a =\,$\,const\,, for instance,\, $\,a =1\,$\,. That is
important difference of\, $\,\widehat{\mathcal{L}}\,\mathcal{V}\,$\,
from $\,\mathcal{\widehat{L}}\,\mathcal{F}\,$. Consequently, taking
in mind the action of $\,\widehat{\mathcal{L}}\,$ onto
$\,\mathcal{V}\,$, one can write
\begin{equation}
\mathcal{\widehat{L}}\left(\sigma +\psi\,,\frac {\delta }{\delta
\psi}\right )\,=\,\mathcal{\widehat{L}}\left(\frac
{\psi}{1+\sigma}\,,\frac {\delta }{\delta\, [\,\psi/(1+\sigma
)\,]}\right )\,\,\label{lt}
\end{equation}
At last, let us look at the operator
$\,\mathcal{\widehat{L}}^{\,\,\prime}\,$ (the formula next to
(\ref{fev})). In contrast to $\,\mathcal{\widehat{L}}\,$, it depends
on the density $\,\nu\,$. Nevertheless, with the help of equality
(\ref{eqid}) it is easy to make sure that it obeys a relation like
(\ref{lt}) if the transformation of argument $\,\psi({\bf r},{\bf
p})\,$ is accompanied by transformation of argument $\,\nu\,$ in
accordance with (\ref{nd}):
\begin{equation}
\mathcal{\widehat{L}}^{\,\,\prime}\left(\nu\,,
\,\sigma +\psi\,,\frac
{\delta }{\delta \psi}\right
)\,=\,\mathcal{\widehat{L}}^{\,\,\prime}
\left(\,\upsilon(\sigma,\nu)\,,\,\frac
{\psi}{1+\sigma}\,,\frac {\delta }{\delta\, [\,\psi/(1+\sigma
)\,]}\right )\,\,\label{lpt}
\end{equation}

The combination of all these observations implies the invariance
property
\begin{equation}
\mathcal{V}\{t,{\bf R},{\bf P},\,\sigma +\psi\,|\,{\bf
R}_0;\,\nu\}\,=\,\mathcal{V}\left\{t,{\bf R},{\bf P},\,\frac
{\psi}{1+\sigma}\,|\,{\bf R}_0;\,\upsilon (\sigma
,\nu)\right\}\,\,\,\label{id}
\end{equation}
or, equivalently and similarly to (\ref{gr}),
\begin{eqnarray}
\widehat{\mathcal{T}}(\sigma)\,\mathcal{V}\{t,{\bf R},{\bf
P},\,\psi\,|\,{\bf R}_0;\,\nu\}\,\equiv\,\mathcal{V}\{t,{\bf R},{\bf
P},\,\frac {1+\psi}{1+\sigma}-1\,|\,{\bf R}_0;\,\upsilon (\sigma
,\nu)\}\,=\,\nonumber\\
=\,\mathcal{V}\{t,{\bf R},{\bf P},\,\psi\,|\,{\bf
R}_0;\,\nu\}\,\,\,,\,\,\,\,\,\,\,\,\, \,\,\,\, \,\,\,\, \,\,\,\,
\,\,\,\, \,\,\,\, \,\,\,\, \,\,\,\, \,\,\,\,\label{grv}
\end{eqnarray}
where the left equality, together with (\ref{ccc}), (\ref{nd}),
(\ref{gr}) and (\ref{gr1}), defines action of the above characterized
group of transformations onto functional $\,\mathcal{V}\{t,{\bf
R},{\bf P},\psi\,|\,{\bf R}_0;\,\nu\}\,$. Apparently, in respect to
this functional that are such the transformations of its arguments
which do not change its value. The expansion of the equality
(\ref{id}) into series in terms of $\,\psi\,$ yields
\begin{eqnarray}
V_0(t,{\bf R},{\bf P}|\,{\bf R}_0;\,\upsilon (\sigma ,\nu))\,
=\,V_0(t,{\bf R},{\bf P}|\,{\bf R}_0;\,\nu)\,+\,\,\,\,\,\,\,
\,\,\,\,\,\,\,\,\,\,\,\,\, \,\,\,\,\,\,\,\,\,\,\,\,\,\,\,\,\,
\,\,\,\,\,\,\,\,\,\,\,\,\, \,\,\,\,\,\,\,\,\,\,\,\, \,\,\, \label{vexp}\\
\,\,\,\,\,\,\,\,\,\,\,\, \,\,\,\,\,\,\,\,\,\,\,\,\,\,\,\,\,\,\,\,
\,\,\,\,\,\,\,\,\, \,\,\, \,\,\,\,\,\,\,\, \,\,\,\,\,\,\,\,\
\,\,\,\,\,\,\,\,\,\,\,\,\ +\,\sum_{n\,=\,1}^\infty \frac
{\nu^n\sigma^n}{n!}\,\int_1\! ...\!\int_n V_n(t,{\bf R},{\bf
r}^{(n)},{\bf P},{\bf p}^{(n)}|\,{\bf R}_0;\,\nu)\,\,\,,\nonumber\\
\left[\frac {\upsilon (\sigma ,\nu)}
{(1+\sigma)\nu}\right]^k\,V_k(t,{\bf R},{\bf r}^{(k)},{\bf P},{\bf
p}^{(k)}|{\bf R}_0; \,\upsilon (\sigma ,\nu))\,=\, V_k(t,{\bf R},{\bf
r}^{(k)},{\bf P},{\bf p}^{(k)}|{\bf
R}_0;\,\nu)\,+\nonumber\\
+\sum_{n\,=\,1}^\infty \frac {\nu^n\sigma^n}{n!}\int_{k+1}\! ...
\!\int_{k+n} V_{k+n}(t,{\bf R}, {\bf r}^{(k+n)},{\bf P},{\bf
p}^{(k+n)}|{\bf R}_0;\,\nu)\,\,\label{vexpn}
\end{eqnarray}
Corresponding infinitesimal (in respect to $\,\sigma\,$)
relations are similar to (\ref{eqinf})\,:
\begin{eqnarray}
\left\{n\,\varkappa(\nu)\,+\,[\,1+\varkappa(\nu)\,] \,\nu\,\frac
{\partial}{\partial \nu}\,\right\} \,V_n(t,{\bf R},{\bf r}^{(n)},{\bf
P},{\bf p}^{(n)}|{\bf R}_0;\,\nu)\,=\,\,\, \,\,\,\,\,
\,\,\,\,\,\,\,\,\,\,\,\,\,\label{inf}\\ \,\,\,\,\,\,\,\, \,=
\,\nu\int_{n+1} V_{n+1}(t,{\bf R},{\bf r}^{(n+1)},{\bf P},{\bf
p}^{(n+1)}|{\bf R}_0;\,\nu) \,\nonumber
\end{eqnarray}

Formulas (\ref{vexp}) and (\ref{vexpn}) can be interpreted as
``virial expansions'' of the probabilistic law of BP's random
wandering and statistical correlations between BP and medium, with
those difference from the usual virial expansions of thermodynamic
quantities \cite{ll1} or kinetic coefficients \cite{ll2} that here
some decrements of the density do appear instead of its full value.
However, in the limit when\, $\,\nu\rightarrow 0\,$ and
$\,\sigma\rightarrow \infty\,$\, with\, $\,\nu\sigma =\,$const\,\,
our relations take quite usual form. For instance, (\ref{vexp})
transforms into
\begin{eqnarray}
V_0(t,{\bf R},{\bf P}|{\bf R}_0;\upsilon (\nu)) =V_0(t,{\bf R},{\bf
P}|{\bf R}_0;0)+\sum_{n\,=\,1}^\infty \frac {\nu^n}{n!}\int
V_n(t,{\bf R},{\bf r}^{(n)},{\bf P},{\bf p}^{(n)}|{\bf
R}_0;0)\,\,,\label{sexp}\\
\upsilon(\nu)\,\equiv\,\nu \left[1+\sum_{n\,=\,1}^{\infty }\frac
{\nu^n}{n!}\int_1\! ...\!\int_n C_{n+1}({\bf r},{\bf r}_1\,...\,{\bf
r}_n;0)\right]\,\,\,,\,\,\,\,\, \,\,\,\,\,\,\,\,\, \,\,\,\,\,\,\,
\nonumber
\end{eqnarray}
where\, $\,\int\,$ in the first row means integration over all the
atoms' variables. It should be noted that formula (\ref{vexp}) was
obtained earlier in \cite{mpa3,pro,jstat} on the base of the
``generalized fluctuation-dissipation relations''  \cite{fds,p}. Up
to that time, as far as I know, such kind of exact relations was not
in use in statistical physics. It would be rather hard to extract
them directly from the BBGKY hierarchy (\ref{fn}) or equivalent
equations for CF. The only exclusion is the special case of BP in
ideal gas \cite{last,py}, when\, $\,C_n=0\,$\,, and readers of this
paper can easy derive relations (\ref{inf}) (with
$\,\varkappa(\nu)=0\,$) by means of differentiation of equations
(\ref{vn}) in respect to gas density.

\vspace{-10 pt}
\section{On principal consequences from the invariance relations}
An exact solution of the BBGKY hierarchy automatically satisfies all
the exact ``virial expansions'' (\ref{vexp})-(\ref{sexp}) resulting
from the above-stated invariance group. Moreover, expansion
(\ref{sexp}) itself gives explicit formal solution to the BBGKY
hierarchy (since at $\,\nu =0\,$ the corresponding equations for CF
can be solved by means of direct successive quadratures, like
equations (\ref{vn}) at $\,\nu =0\,$). Therefore it is natural to
apply our above results to ``testing of statistical hypotheses''
about the BP's wandering.

In this respect it is important to emphasize that formulas
(\ref{vexp})-(\ref{vexpn}) interconnect two different random walks in
two media with arbitrarily assigned densities ratio,\, $\,\upsilon
(\sigma ,\nu)/\nu\,$. Consequently, all the terms in series
(\ref{vexp}) or (\ref{vexpn}), as well as in (\ref{sexp}), are
equally necessary for correct summation of the series, and its
truncation would produce not an approximative but wrong result. Even
in the ``dilute gas limit'' (when gas parameters tend to zero,
$\,4\pi r_a^3\nu/3\rightarrow 0\,$, $\,4\pi r_b^3\nu/3\rightarrow
0\,$ with\, $\,r_{a,\,b}\,$\, being characteristic radii if atom-atom
and BP-atom interactions) or in the so called ``Boltzmann-Grad limit'
(when simultaneously the mean free paths,\, $\,\Lambda_b= (\pi
r_b^2\nu)^{-1}\,$ and $\,\Lambda_a= (\pi r_a^2\nu)^{-1}\,$,\, are
fixed), as well as in the case of ideal gas (when
$\,\Lambda_a=\infty\,$ but $\,\Lambda_b\,$ is finite). In all these
cases, formulas (\ref{vexp})-(\ref{inf}) reduce to
\begin{eqnarray}
\frac {\partial^{\,k} V_n(t,{\bf R},{\bf r}^{(n)},{\bf P},{\bf
p}^{(n)}|{\bf R}_0;\nu)}{\partial\nu^{\,k}}=\!\int_{n+1}\! ...
\!\int_{n+k} V_{n+k}(t,{\bf R}, {\bf r}^{(n+k)},{\bf P},{\bf
p}^{(n+k)}|{\bf R}_0;\nu)\,\,,\label{lm}
\end{eqnarray}
thus making it quite obvious that from the point of view of rigorous
theory all the historical correlations are equally substantial.

Meanwhile, it is the convention in physical literature to reject
third- and higher-order inter-particle correlations when deriving
closed equations (kinetic equations) for one-particle DF (for
examples, see \cite{bog,sil,ll2,re}). Sometimes, as in the case of
hard sphere gas \cite{re,vblls}, these correlations get lost even
insensibly for authors (see \cite{last} about that). By the widely
accepted opinion, under the Boltzmann-Grad limit the theory based on
the BBGKY hierarchy should reduce to the classical kinetic theory of
gases (where the BP's probability distribution, $\,F_0=V_0\,$, would
obey either the linearized Boltzmann equation or similar
Boltzmann-Lorentz equation \cite{re,vblls,p1}). However, really from
the aforesaid it follows that such is not the case: the classical
kinetics is not an extreme case of the statistical-mechanical theory.

As an illustration, consider the BP in ideal gas. If rejecting all
the higher-order correlations, i.e. putting on\, $\,V_2=0\,$ in the
second of equations (\ref{vn}) (with $\,n=1\,$) and then inserting
its quadrature into the first equation, we come to a closed equation
for\, $\,V_0\,$\,. An asymptotic of solution of this equation at
$\,t\gg \tau =\Lambda_b /v_0 \,$\, (where\, $\,v_0\sim\sqrt{T/M}\,$
is characteristic thermal velocity of BP) is certainly Gaussian:
\begin{equation}
V_0(t,\Delta{\bf R};\nu)\equiv\!\int V_0(t,{\bf R},{\bf P}|{\bf
R}_0;\nu)\,d{\bf P}\,\rightarrow\,\frac {\exp{(-\Delta {\bf
R}^2/4Dt)}}{(4\pi Dt)^{\,3/2}}\,\,\,, \label{vg}
\end{equation}
where\, $\,\Delta {\bf R}={\bf R}-{\bf R}_0\,$\,, and \,$\,D\sim
v_0\Lambda_b\,$ is the BP's diffusivity. For a rarefied gas,\,
$\,\Lambda_b=(\pi r_b^2\nu)^{-1}\,$\, and \,$\,D\propto
\nu^{-1}\,$\,. Anyway, (\ref{vg}) is some rather complicated function
of\, $\,\nu\,$\,. At the same time, from the point of view of the
exact relations (\ref{lm}), the equality $\,V_2=0\,$ immediately
implies that $\,V_0(t,\Delta{\bf R};\nu)\,$ is purely linear function
of\, $\,\nu\,$ !

So strong discrepancy does mean that asymptotic (\ref{vg}) is
incorrect, and we have to get back to the complete BBGKY hierarchy.
An approximate approach to its correct analysis was suggested already
in \cite{i1} (or see \cite{i2}) and developed in \cite{p1}. In
\cite{p1}, for particular situation when the role of BP is played by
a marked gas atom, the following asymptotic was found:
\begin{equation}
V_0(t,\Delta {\bf R};\,\nu)\,\rightarrow \,\,\frac {\Gamma
(7/2)}{[4\pi D\, t\,]^{\,3/2}}\, \left[\,1+\frac {\Delta{\bf
R}^2}{4D\,t}\,\right]^{-\,7/2}\Theta \left (\frac {|\Delta {\bf
R}|}{v_0 t}\right )\,\,\label{as}
\end{equation}
Here,\, $\,\Theta(x)\,$ is definite function which turns to unit at
$\,x=0\,$ and rapidly goes to zero at $\,x\rightarrow\infty\,$. The
characteristic feature of this probability distribution is that its
fourth-order cumulants grow with time not linearly (as it would be in
the case of Gaussian asymptotic) but under the law\, $\,\propto
(Dt)^2\,\ln (t/\tau )\,$\,, that is nearly proportionally to the
square of time.

Empirically, such the statistical property of the random walk is
taken for the fluctuations in current diffusivity of BP
\cite{bk3,i1,bk12} whose power spectral density at frequencies $\,f
\ll 1/\tau\,$ is approximately inversely proportional to the
frequency (so called ``1/f\,-noise'' \cite{bk3};\, thus, results of
\cite{i1,p1} confirmed the conjectures about origin of 1/f\,-noise
stated for the first time in \cite{bk12}). From the physical
viewpoint, the origin of these fluctuations and corresponding
``historical'' statistical correlations described by CF $\,V_{n}\,$
($\,n>0\,$) is mere indifference of the system to the pre-history of
BP's collisions, i.e. to their total number and their time-average
rate, as well as to ratios of numbers of collisions with different
impact parameters \cite{i1}. Thus, the historical correlations are
caused by complicity of particles in their collision rate
fluctuations rather than by their interactions as such.

Let us demonstrate that quite similar conclusions can be deduced
basing on only the virial expansions and the trivial non-negativeness
of all the DF $\,F_n\,$.

From the non-negativity of $\,F_1\,$ and identity (\ref{cf1}) we have
\begin{eqnarray}
V_0(t,\Delta {\bf R};\,\nu)\int_{\Omega} F_1^{(eq)}({\bf r}|{\bf
R};\nu)\,d{\bf r}\,+\int_{\Omega}\,V_1(t,{\bf R},{\bf r}|{\bf
R}_0;\,\nu)\,d{\bf r} \,\geq\,0\,\,\,,\label{in0}
\end{eqnarray}
where\,\, $\,V_1(t,{\bf R},{\bf r}|{\bf R}_0;\,\nu)\, \equiv\,\int
\int\, V_1(t,{\bf R},{\bf r},{\bf P},{\bf p}|{\bf R}_0;\,\nu)\,\,
d{\bf p}\,d{\bf P}\,$\,\,, and\, $\,\Omega\,$ is any region in the
space of vectors\,  $\,{\bf r}-{\bf R}\,$\,. At given\, $\,0<\delta
<1\,$,\,  let\, $\,\Omega(\delta,t,\Delta{\bf R};\nu)\equiv
\Omega(\delta)\,$\, be the minimum (in the sense of volume) among all
regions $\,\Omega\,$ which satisfy the condition
\begin{eqnarray}
\left|\int_{\Omega} V_1\,d{\bf r}\,-\,\int V_1\,d{\bf
r}\,\right|\,\leq\,\delta\,\left|\int V_1\,d{\bf
r}\,\right|\,\,\label{df}
\end{eqnarray}
From this definition it follows that
\begin{eqnarray}
\Omega\,\max_{\,{\bf r}} |\,V_1|\, \,\geq\, \left|\,\int_{\Omega}
V_1\,d{\bf r}\,\right|\,\geq\, (1-\delta)\,\left|\,\int V_1\,d{\bf
r}\,\right|\,\,\label{df1}
\end{eqnarray}
(spatial region and its volume are denoted by the same letter). In
the aggregate the inequalities (\ref{in0}) and (\ref{df}) produce, as
can be easy verified, inequality
\begin{equation}
V_0(t,\Delta {\bf R};\,\nu)\int_{\Omega(\delta)} F_1^{(eq)}({\bf
r}|{\bf R};\nu)\,d{\bf r}\,+\,(1-\delta)\int V_1(t,{\bf R},{\bf
r}|{\bf R}_0;\,\nu)\,d{\bf r}\,\,\geq\,\,0\,\,\label{in}
\end{equation}
Combining it with the first of the virial expansions (\ref{inf})
(with $\,n=0\,$) and taking into account that in the case under
consideration $\,\varkappa(\nu)\rightarrow 0\,$\, and $\,
F_1^{(eq)}({\bf r}|\,{\bf R};\nu)\leq 1\,$, after quite obvious
reasonings one obtains
\begin{equation}
\overline{\Omega}(t,\Delta {\bf R};\,\nu)\,V_0(t,\Delta {\bf
R};\,\nu)\,+\,\frac {\partial V_0(t,\Delta {\bf R};\,\nu)}{\partial
\nu}\,\, \geq\,0\,\,\,, \label{in1}
\end{equation}
where the quantity\, $\,\overline{\Omega}(t,\Delta {\bf R};\,\nu)\,$
if defined as follows:
\begin{equation}
\overline{\Omega}(t,\Delta {\bf R};\,\nu)\,=\,\min_{0<\,\delta
<1}{\frac {\Omega(\delta,t,\Delta{\bf R};\nu)}{1-\delta}}\,\,
\label{md}
\end{equation}

Next, discuss a meaning of the latter quantity. Notice, first, that\,
$\,\Omega(\delta)\rightarrow 0\,$ when $\,\delta \rightarrow 1\,$,
merely by the definition of $\,\Omega(\delta,t,\Delta{\bf
R};\nu)=\Omega(\delta)\,$. Second, if the function\, $\,V_1(t,{\bf
R},{\bf r}|{\bf R}_0;\,\nu)\,$ does not change its sign when changing
argument $\,{\bf r}-{\bf R}\,$\, (but keeping other arguments fixed)
then at $\,\delta \rightarrow 1\,$ the region $\,\Omega(\delta)\,$
shrinks into infinitely small neighborhood of the point of maximum of
$\,|V_1|\,$ (or union of such neighborhoods), so that one can write
\begin{eqnarray}
\left|\,\int_{\Omega(\delta)} V_1\,d{\bf r}\,-\,\int V_1\,d{\bf
r}\,\right|\,\rightarrow\,\left|\,\int V_1\,d{\bf
r}\,\right|\,-\,\Omega(\delta)\,\max_{\,{\bf r}} |\,V_1|\,\,\,, \nonumber\\
\,\,\,\Omega(\delta)\,\max_{\,{\bf r}} |\,V_1|\,\rightarrow\,
(1-\delta)\, \left|\,\int V_1\,d{\bf r}\,\right|\,\nonumber
\end{eqnarray}
Consequently,\, $\,\overline{\Omega}(t,\Delta {\bf R};\,\nu)\,\leq
\,|\,\int V_1\,d{\bf r}\,|/(\max_{\,{\bf r}} |\,V_1|)\,$.\, From the
other hand, because of the inequality (\ref{df1}) definitely\,
$\,\overline{\Omega}(t,\Delta {\bf R};\,\nu)\,\geq \,|\,\int
V_1\,d{\bf r}\,|/(\max_{\,{\bf r}} |\,V_1|)\,$. From here (recalling
the constancy of $\,V_1$'s sign) we conclude that
\begin{equation}
\overline{\Omega}(t,\Delta {\bf R};\,\nu)\,=\,\frac {\int
|\,V_1(t,{\bf R},{\bf r}|{\bf R}_0;\,\nu)|\,d{\bf r}}{\,\max_{\,{\bf
r}} |\,V_1(t,{\bf R},{\bf r}|{\bf R}_0;\,\nu)|\,}\,\, \label{md1}
\end{equation}
This is nothing but the effective volume of a region in $\,({\bf
r}-{\bf R})$-space on which the two-particle correlation spreads out.
Now, let our assumption about the sign constancy of\, $\,V_1(t,{\bf
R},{\bf r}|{\bf R}_0;\,\nu)\,$ in respect to $\,{\bf r}-{\bf R}\,$\,
is not satisfied, and $\,\mathcal{D}\,$ denote those part of the
$\,({\bf r}-{\bf R})$-space where $\,V_1\,$ has dominating sign, that
is coinciding with sign of integral $\,\int V_1\,d{\bf r}\,$. Then
the analogous reasonings yield
\begin{equation}
\frac {\left|\,\int V_1\,d{\bf r}\,\right|}{\,\max_{\,{\bf r}}
|\,V_1|\,}\,\leq\,\overline{\Omega}(t,\Delta {\bf
R};\,\nu)\,\leq\,\frac {\left|\,\int V_1\,d{\bf
r}\,\right|}{\,\max_{\,\,{\bf r}\in\, \mathcal{D}}
|\,V_1|\,}\,\leq\,\frac {\int_{\mathcal{D}} |\,V_1|\,d{\bf
r}}{\,\max_{\,\,{\bf r}\in\, \mathcal{D}} |\,V_1|\,}\,\, \label{md2}
\end{equation}
Hence, in such the case we can say that $\,\overline{\Omega}(t,\Delta
{\bf R};\,\nu)\,$ is even less than effective volume of a region
where significant two-particle correlation takes place.

Further, let us confront inequality (\ref{in1}) with the hypothesis
that the asymptotic of $\,V_0(t,\Delta {\bf R};\,\nu)\,$ has Gaussian
shape (\ref{vg}). From (\ref{in1}) it follows that this hypothesis
can be true only if
\begin{equation}
\overline{\Omega}(t,\Delta {\bf R};\,\nu)\,\geq\,-\,\frac
{\partial\ln D}{\partial \nu}\left(\frac {\Delta {\bf
R}^2}{4Dt}-\frac 32\right)\,=\,\frac {1}{\nu}\,\left(\frac {\Delta
{\bf R}^2}{4Dt}-\frac 32\right)\,\, \label{a}
\end{equation}
(we took into account that in gas\, $\,D\propto \nu^{-1}\,$). In
other words, if the mentioned effective volume, occupied by the
two-particle correlation, is not bounded above under variations of\,
$\,t\,$ and $\,\Delta {\bf R}\,$. In opposite, if this volume is
bounded above, and
\begin{equation}
\nu\,\overline{\Omega}(t,\Delta {\bf
R};\,\nu)\,\leq\,c_1\,=\,\mathrm{const}\,\,\,, \label{bound}
\end{equation}
then the Gaussian asymptotic (\ref{vg}) is forbidden. Instead of it,
inequality (\ref{in1}) allows the asymptotic like
\begin{equation}
V_0(t,\Delta {\bf R};\,\nu)\,\rightarrow\,\frac {1}{(4\pi
Dt)^{3/2}}\,\,\Psi\left(\frac {\Delta {\bf R}^2}{4Dt}\right)\,
\Theta\left(\frac {|\Delta {\bf R}|}{v_0t}\right) \,\,\,, \label{as1}
\end{equation}
where function $\,\Psi(z)\,$ should satisfy the inequality
\begin{equation}
z\,\frac {d\Psi(z)}{dz}\,+\,\alpha\,\Psi(z)\,\geq\,0\,\,\,,
\,\,\,\,\,\,\,\alpha\,\equiv\,\frac 32\,+\,c_1\left[-\,\frac
{\partial\ln D}{\partial \ln\nu}\right]^{-1}\,=\,\frac 32\,+\,c_1
\,\label{in2}
\end{equation}
Consequently, $\,\Psi(z)\,$ can not decrease at infinity faster than
under the power law:\,
$\,\Psi(z\rightarrow\infty)\,\propto\,1/z^{\,\alpha}\,$\,. The
formula (\ref{as}) in \cite{p1} corresponds to
$\,c_1=\nu\overline{\Omega}=2\,$.

Thus, the theory inevitably leads to the statistical correlations
whose length appears unbounded if not in space, as at variant
(\ref{a}), then in time, as at variant (\ref{bound}). It remains to
ascertain what of these variants is more close to exact solution of
the BBGKY hierarchy. From the physical viewpoint, the second is
undoubtedly preferable. Indeed, according to equations (\ref{vn}),
the source of the correlations between BP and atoms is their
collisions. A collision intends such relative disposition of BP and
an atom when the vector  $\,\rho={\bf r}-{\bf R}\,$ belongs to the
``collision cylinder'' which is oriented in parallel to the relative
velocity $\,{\bf u}={\bf p}/m-{\bf P}/M\,$ and has radius $\,\approx
r_b\,$. At the same time, the particles should not be separated by a
distance much larger than\, $\,\Lambda = \min
{\,(\Lambda_a,\Lambda_b)}\,$,\, since otherwise their collision most
likely would be prevented by an encounter of one of them with other
gas particles. Hence, the two-particle correlations are concentrated
in the middle of the collision cylinder, at\, $\,-\Lambda\lesssim
{\bf u}\cdot\rho/|{\bf u}| \lesssim\Lambda\,$. The volume of this
space region does not depend on the particles' momenta, therefore it
gives natural estimate of the volume $\,\overline{\Omega}(t,\Delta
{\bf R};\,\nu)\,$. Obviously, it does not depend also on $\,t\,$ and
$\,\Delta {\bf R}\,$. Thus, $\,\overline{\Omega}\,\approx\,
2\Lambda_b\,\pi r_b^2\,=\,2/\nu\,$ (if, like on \cite{p1}, BP is
merely one of atoms). As the result, we come to (\ref{in2}) with
$\,c_1=\nu\overline{\Omega}\approx 2\,$.

\section{Conclusion}
To resume, the problem about thermodynamically equilibrium random
walk of a test ``Brownian'' particle (BP) in a fluid was formulated
as a problem of classical statistical mechanics in the framework of
the Gibbs canonic statistical ensemble of initial conditions. The
corresponding BBGKY equations for distribution functions (DF) and the
Bogolyubov equation for their generating functional were considered
in terms of the cumulant functions (CF) introduced in order to
extract statistical correlations between current state of the fluid
and total displacement, or path, of the BP during all the time of its
observation (``historical correlations''). It is shown that the
generating functionals of equilibrium DF and time-dependent DF both
possess invariance in respect to definite continuous group of
transformations of their arguments, including the density of the
fluid (mean concentration of atoms).

The obtained invariance group produces the exact ``virial
expansions'' (``virial relations'') which connect full sets of CF
taken at different values of the density. With the help of these
relations it was shown that all the many-particle correlations play
equally important roles, regardless of the gas parameter value, and
therefore the standard ``Botzmannian'' approach to kinetics of BP,
which rejects third- and higher-order correlations, is incorrect even
in the low-density or Boltzmann-Grad limit. Correspondingly, the
virial relations quite certainly forbid the Gaussian asymptotic of
the BP's path probability density distribution inherent to the
Boltzmann's kinetics. Instead, such an asymptotic is allowed only
which has power-law long tails (cut off at the BP's path values on
order of the ballistic flight path). This conclusion agrees with the
results of approximative solving of the BBGKY equations \cite{i1,p1}.
It means that all the constituent parts of trajectories (random
walks) of BP are statistically dependent one on another. Therefore,
the trajectories cannot be imitated by sequences of mutually
independent, in the sense of the probability theory, random trials
(speaking figuratively and in the Einstein words,\, ``God does not
play dice''\,). In this sense, any real molecular random walk, in
contract to the random walk in the Lorentz gas, is non-ergodic.

It should be noted that results of the present paper can be easily
extended to thermodynamically non-equilibrium random walks under
influence of an external force (in this respect see
\cite{mpa3,jstat,i2}). It would be interesting also to generalize the
considered invariance group to another problems focused on not
selected particles only but sooner on collective and hydrodynamical
variables.

I would like to acknowledge my colleagues from DonPTI NANU Dr.
I.\,Krasnyuk and Dr. Yu.\,Medvedev for many useful
discussions\,\footnote{\,  I am very grateful also to Reviewer of
this paper from the ``Theoretical and mathematical physics''
(``Teoreticheskaya i matematicheskaya fizika'' where its Russian
original is accepted) for criticism which stimulated its noticeable
improvement.}\,.


\end{document}